\documentclass[12pt,preprint]{aastex}

\usepackage{geometry}      
\usepackage{graphicx}
\usepackage{amssymb}
\usepackage{epstopdf}

\newcommand{\feh}{\hbox{$ [{\rm Fe}/{\rm H}]$ }}

\begin{document}

\title{Distance to the Sagittarius Dwarf Galaxy using MACHO Project RR Lyrae stars}

\author{Andrea Kunder and Brian Chaboyer} 
\affil{Dartmouth College, 6127 Wilder Lab, Hanover, NH 03755}
\affil{E-mail: Andrea.Kunder@Dartmouth.edu and brian.chaboyer@dartmouth.edu}

\begin{abstract}
We derive the distance to the northern extension of the 
Sagittarius (Sgr) dwarf spheroidal galaxy from 203 Sgr RR0 Lyrae stars
found in the MACHO database.  Their distances are determined
differentially with respect to 288 Galactic Bulge RR0 Lyrae stars 
also found in the MACHO data.  We find a distance modulus difference 
of 2.41 mags at $l$ = 5$^{\circ}$ and $b$ = -8$^{\circ}$ and that the 
extension of the Sgr galaxy
towards the galactic plane is inclined toward us.
Assuming $\rm R_{GC}$ = 8 kpc, this implies the distance to 
these stars is $(m-M)_0$ = 16.97 $\pm$ 0.07 mags, which corresponds to 
D = 24.8 $\pm$ 0.8 kpc.
Although this estimate is smaller than previous determinations for this
galaxy and agrees with previous suggestions that Sgr's body is 
truly closer to us, this estimate is larger than studies
at comparable galactic latitudes.
\end{abstract}

\section{Introduction}
In our Milky Way galaxy, RR Lyrae stars have advanced our understanding
of the structures of the halo.  It is clear that the outer regions of
the halo are not a smooth distribution, but quite clumpy, and the 
interpretations suggest that these sub-structures are relics of small
satellite galaxies that have been accreted and destroyed by the
tidal forces of the Milky Way \citep{newberg02, vivas01, yan00}.
In order to model and quantify how important
such interactions are in the formation of the halo, fundamental 
parameters such as the distance to the main body from which the remains of 
the disruption process originate, are needed.  The Sagittarius dwarf 
spheroidal galaxy (Sgr) is a striking example of a nearby satellite galaxy 
of the Milky Way that is currently under the strain of the Galactic tidal field
\citep{ibata94, ibata97, monaco04}.

The Sagittarius dwarf spheroidal galaxy has been the subject
of much debate since its discovery by \citet{ibata94}.
Although the broad consensus is that the Sgr is a tidally
disrupted satellite distributed across much of the celestial sphere,
several major issues remain controversial and intertwined 
\citep[see][]{majewski03}.  Advancements in observational constraints
can greatly improve models for the interaction of Sgr with the
Milky Way and can increase the current understanding of both
the Milky Way and the Sgr Galaxy.

When modeling the structure of the tidal debris, parameters constrained by 
observations of stars associated with Sgr are incorporated, 
\textit{i.e.}, distances, velocities, surface densities.  The model that 
best matches the observational data dictates the estimated mass
and orbit of the Sgr Galaxy.  Although features in the 
observational data have been explained by models of the Sgr stream 
\citep[e.g.,][]{johnston99}, many conclusions are only tentative, because
they rely heavily on the less certain measurements of debris properties.

The absence of data from the Sgr galaxy in important regions of the sky
has also hampered investigations pertaining to the Galactic halo.
For example, \citet{helmi04} provides simulations of the Sgr stream 
for a range of halo shapes from extreme oblate to prolate, all of 
which broadly agree with the data available at that time.

The center of Sgr has been studied more then any other constituent part.
The properties of the debris emanating from the main body are particularily 
uncertain and the most subject to speculation.  For example,
because the Sun lies close enough to Sgr orbital plane to be 
well within the width of the Sgr tidal debris stream \citep{majewski03}, 
there may be Sgr debris close to us.  
But this is dependent on where the debris crosses the Galactic plane
on this side of the Galactic Center and on the length of the leading arm.
Some models (e.g., Ibata et al. 2001) derive 
orbits for Sgr that predict current passage of leading 
arm debris through the Galactic plane at a mean distance of $\sim$ 4 kpc 
outside the Solar Circle, while other models obtain a passage of the center of 
the leading Sgr arm debris within two kilo-parsecs of the Sun 
\citep{majewski03}.  \citet{newberg07} use SDSS photometry 
of blue horizontal-branch and F turnoff stars to extrapolate the path 
of the Sgr leading tidal tail and find that it misses the Sun by more 
than 15 kpc.

The tidal debris in the Sgr neighborhood is beginning to be traced out.
Probably the most complete picture of the Sgr stream was
obtained by \citet{majewski03} using M giants selected from the
Two Micron All-Sky Survey (2MASS).  
They could trace out the Sgr leading tidal tail reaching toward the 
North Galactic Cap and the trailing tidal tail in the Southern
Galactic hemisphere.  Recently, \citet{belokurov06} saw the continuation
of the leading tidal trail through the Galactic Cap and into
the Galactic Plane.  One key in addressing questions about the 
orbital path of Sgr
is to determine distances along the stream, and to better
define the projected distribution of Sgr stars on the sky.

Studies of RR Lyrae stars have been instrumental in acquiring data
of the more obscured regions in the leading tidal tail close
to the Galactic Plane (Cseresnjes, Alard \& Guibert 2000; Alard 1996;
Alcock et al. 1997).
Databases like that of MACHO allow for the study of nearby galaxies,
such as the Sgr dwarf galaxy, located behind the Galactic bulge.
\citet{alc97} were the first to use RR0 Lyrae stars in the 
MACHO database to derive a distance to the Sgr Dwarf Galaxy.
Their analysis is restricted in many ways, particularly since it was based
on $\sim$ 24 Sgr RR0 Lyrae stars.
This paper provides a robust distance estimate to the Sgr galaxy
using $\sim$ 200 Sgr RR Lyrae stars from the MACHO database.  The procedure
used here carefully minimizes systematic and statistical errors
and leads to a distance estimate with the smallest formal error to
date.

The models of Sgr already generated in the literature demonstrate an 
immense potential for using debris to determine 
Sgr's dynamical history in great detail.  The accurate distance estimate
to the northern extension of the Sgr galaxy (in Galactic coordinates)
presented here is an important step in constraining Sgr models.

\section{Data and Photometry}
The MACHO Project data collection and experiment are described by
\citet{alc96} and was designed to search for gravitational microlensing events.
Through the simultaneous imaging of two-color photometry on millions of 
stars in the LMC, SMC, and Galactic bulge from 1992 to 1999, many
variable stars were also found.
This paper uses the RR0 Lyrae stars from the MACHO Bulge fields 
\citet{kunder08a} with photometry calibrated to Johnson $V$ and 
Kron-Cousins $R$ bandpasses following \citet{alc99}.

It has been noted that because of the non-standard passbands, the severe 
"blending'' problems in the fields close to the Galactic Bulge, and the 
complexity of the calibration procedures, 
the absolute photometric calibration of the MACHO variable stars is a concern.
With a microlensing search, only differential photometry is needed;
a transformation to the standard system and individual field zero-points
are not priority tasks for the survey telescope.
Because of these photometry difficulties and in order
to avoid systematic effects, the analysis here will be restricted to a 
differential
approach administered on a field by field basis ({\it i.e.}, determining
relative distances between the bulge and Sgr in each MACHO field).

An internal precision of $\sigma_V=0.021$ mag (based on 20,000 stars with 
$V \la 18$ mag) is quoted by \citet{alc99}.  The Sgr stars, however, 
have $V$ magnitudes greater than 18 mag.  To determine the
internal precision of $V >$ 18 mag, a Fourier decomposition is 
performed on the Bulge and Sgr RR0 Lyrae lightcurves.  The amount 
that each point in the lightcurve deviates from the fit, $\Delta V_{lc}$, 
is then calculated.  Each lightcurve has between 20-700 data points.
The dispersion in the average $\Delta V_{lc}$ will give an indication
of the internal precision.  

The average dispersion in the bulge $< \Delta V_{lc} >$ is 0.06 mags 
(based on 613 representative bulge RR0 Lyrae stars), where the average 
$V$-band magnitude is 16.55 $\pm$ 0.47 mags.
The average dispersion in the Sgr $< \Delta V_{lc} >$ is 0.08 mags 
(based on 175 representative Sgr RR0 Lyrae stars), where the average
$V$-band magnitude is 18.78 $\pm$ 0.28 mags.  A visual 
inspection of the lightcurves suggests that the reason for a dispersion in
$< \Delta V_{lc} >$ that is larger than the published value of the internal
precision of 0.021, is due to a handful largely discrepant points in
the RR0 Lyrae lightcurve that
contribute significantly to the dispersion in $< \Delta V_{lc} > $.  

Removing points with $ \Delta V_{lc} < 0.1$, the average dispersion in the 
bulge $< \Delta V_{lc} >$ is 0.03 mags. On average, 6 points per light curve
were removed, and the number of photometric measurements in each lightcurve 
ranges from 18 to 677 points. For the Sgr sample, the average dispersion 
in the Sgr $< \Delta V_{lc} >$ is 0.04 mags.  The average number
of points removed per lightcurve is also six, and the Sgr lightcurves 
consist of between 56 to 333 measurements. 
Comparing the dispersion of $\Delta V_{lc}$ for the Bulge and Sgr
sample, we can conclude that the Sgr internal precision for the MACHO
fields is about 1.5 times as great as that of the MACHO Bulge fields
with $V < $18 mag.

\section{The RR0 sample}
\subsection{Completeness}
The MACHO RR0 Lyrae sample from \citet{kunder08b} does not have a completeness
estimate.  Their sample was not intended to be a comprehensive MACHO
bulge RR0 Lyrae sample, but rather a representative sample with
well-culled and unambiguous RR0 Lyrae stars.  Here the
completeness of the sample is investigated with particular emphasis on
the completeness as a function of Sgr RR0 Lyrae magnitude.

The two fields of \citet{cseresnjes00} overlap with some of 
the MACHO fields.  This allows an independent check on the approximate 
completeness of the \citet{kunder08b} sample.  Field 2 of the 
\citet{cseresnjes00} data was first processed and presented by 
\citet{alard96}.  They estimated a completeness limit of $B_j$ $=$ 20.1 mags, 
which corresponds to a distance modulus to 18 mags (40 kpc); this limit
was based on the very numerous ($\sim$7000) contact binaries present in the
photographic plates.

\begin{figure}[htb]
\includegraphics[width=16cm]{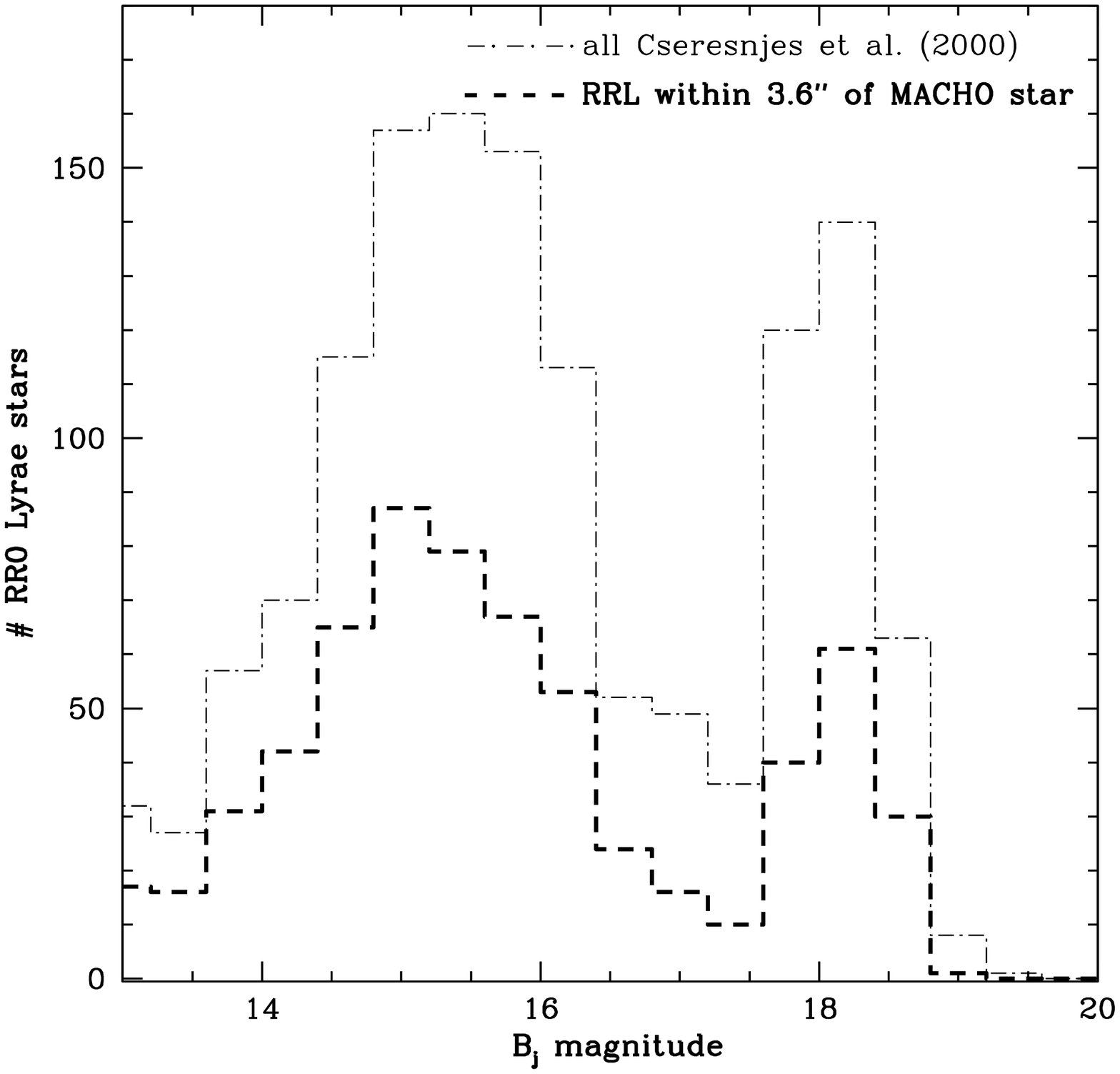}
\caption{Histograms of the MACHO $V$ and \citet{cseresnjes00} $Bj$ magnitudes
of the 675 RR0 Lyrae stars within 3.6 arc-seconds of each other.
A light dashed line shows all the 1353 Field 2 \citet{cseresnjes00} RR0 
Lyrae stars.  The \citet{cseresnjes00} stars matched with the MACHO RR0 Lyrae
stars follow the same distribution as the complete Field 2 sample, and
indicates that the Sgr MACHO RR0 Lyrae
sample is not magnitude limited.  
\label{plotone}}
\end{figure}

Figure~\ref{plotone} shows a histogram of the magnitudes of all 
675 MACHO RR0 Lyrae stars that are within 3.6 arc-seconds of one of the
Field 2 \citet{cseresnjes00} RR0 Lyrae stars.  It is immediately obvious that
the \citet{cseresnjes00} stars matched with the MACHO RR0 Lyrae
stars follow the same distribution as the complete Field 2 sample.  
As the \citet{cseresnjes00} Sgr survey probes much deeper 
than the stars belonging to the Sgr galaxy, this constitutes evidence that 
the Sgr MACHO RR0 Lyrae sample is not magnitude limited to V$\sim$19.5.
The fraction of MACHO Sgr RR0 Lyrae stars that can be matched with 
a \citet{cseresnjes00} RR0 Lyrae star within the magnitude range of 
$B_j$ 15-16 mags, is 50\%.  This drops slightly to 47\% within the $B_j$ 16-17 
magnitude range, to 34\% within the $B_j$ 17-18 magnitude range,
and to 45\% within the $B_j$ 18-19 magnitude range.  This suggests that 
the MACHO RR0 Lyrae sample is at most marginally magnitude limited.  
The reason for the lower fraction of MACHO and \citet{cseresnjes00} 
Sgr RR0 Lyrae stars within the $B_j$ magnitude range that encompasses 
the transition area ($B_j$ 17-18 mag) between the Galactic bulge and 
Sgr galaxy is unclear and could be due to an effect not associated
with magnitude ({\it i.e.}, latitude) or small number statistics.  
If indeed
we assume that the Sgr RR0 Lyrae population contains 5\% less stars 
than the complete sample, then a total of 16 stars, or 10\% of the Sgr
sample is missing due to magnitude limits of the MACHO survey.


The MACHO bulge fields barely reach the low
galactic latitudes of \citet{cseresnjes00} Field 1.  However, they overlap
in a 15$^{\circ}$ x 2.4$^{\circ}$ area.  Between a right ascension of 
18.53 to 18.61h and a declination -29.4 to -27.0$^{\circ}$ there are 145 MACHO
RR0 Lyrae stars and 191 \citet{cseresnjes00} stars.
This field is reported to have a $\sim$ 90\% extraction completeness and 
a $\sim$ 93\% selection completeness, making the MACHO data in this
region 64\% complete.

The MACHO bulge fields cover the majority of \citet{cseresnjes00} Field 2.
Between the right ascension of 18.15 to 18.51h and the declination
of -31.04$^{\circ}$ to -27.1$^{\circ}$, there are 1069 MACHO RR0 Lyrae 
stars and 982
\citet{cseresnjes00} stars.  Their Field 2 has a $\sim$ 70 \% extraction 
completeness and a $\sim$ 85\% selection completeness, making the
MACHO data $\sim$77\% complete in this region.  

From the above analysis, the MACHO RR0 Lyrae sample used by \citet{kunder08b}
is roughly 65\% complete.  More importantly, SGR RR0 Lyrae population is not 
magnitude limited to at least $V \sim$ 20 mag.



\subsection{Absolute Magnitude}
The most popular approach to estimate the RR Lyrae distances is 
a linear $\rm M_{V_{RR}} -[Fe/H]$ relation (e.g., Krauss \& Chaboyer 2003).  
Recently Bono, Caputo, 
\& di Criscienzo (2007) have shown that this relation is
not suitable for the most metal-rich $(\rm \feh > -0.7$ dex)
field variables, and further show that over the metallicity 
range $\rm-2.4<[Fe/H]<0.0$ 
the $\rm M_V(RR)-[Fe/H]$ relation is not linear but has a parabolic behavior:
\begin{equation}
M_V = 1.19 + 0.5\feh + 0.09\feh^2
\label{mvmv}
\end{equation}
A number of studies have shown that Fourier parameters of light curves 
of RR0 Lyrae stars can be used to find their metallicity with an error 
of $\sim$ 0.2 dex \citep[e.g.,][]{jk96, simon93}.  Employing this
technique, \citet{kunder08a} find that the Bulge RR0 Lyrae stars are 
on average $\sim$ 0.28 $\pm$ 0.02 dex more metal rich then the average Sgr RR0 
Lyrae in the MACHO bulge fields, with $\feh_{Sgr}$=$-$1.55 dex.  
This corresponds to a $\sim$ 0.15 mag
offset in absolute magnitude, which at the distance of Sgr translates 
into a $\sim$ 1.7 kpc error in the distance.  In the paper we use
the RR0 Lyrae stars with \feh metallicities derived from 
\citet{kunder08a} so that the metallicity dependence of the absolute
magnitude in the RR Lyrae stars can be taken into account.
The inclusion of the RR Lyrae stars metallicity dependence on its absolute
magnitude, is in contrast to most previous Sgr distance estimates {\it e.g.},
\citet{mateo95,alard96,cseresnjes00} which all assume a constant 
$M_{V_{RR}}$.

\subsection{Distribution}
\begin{figure}[htb]
\includegraphics[width=16cm]{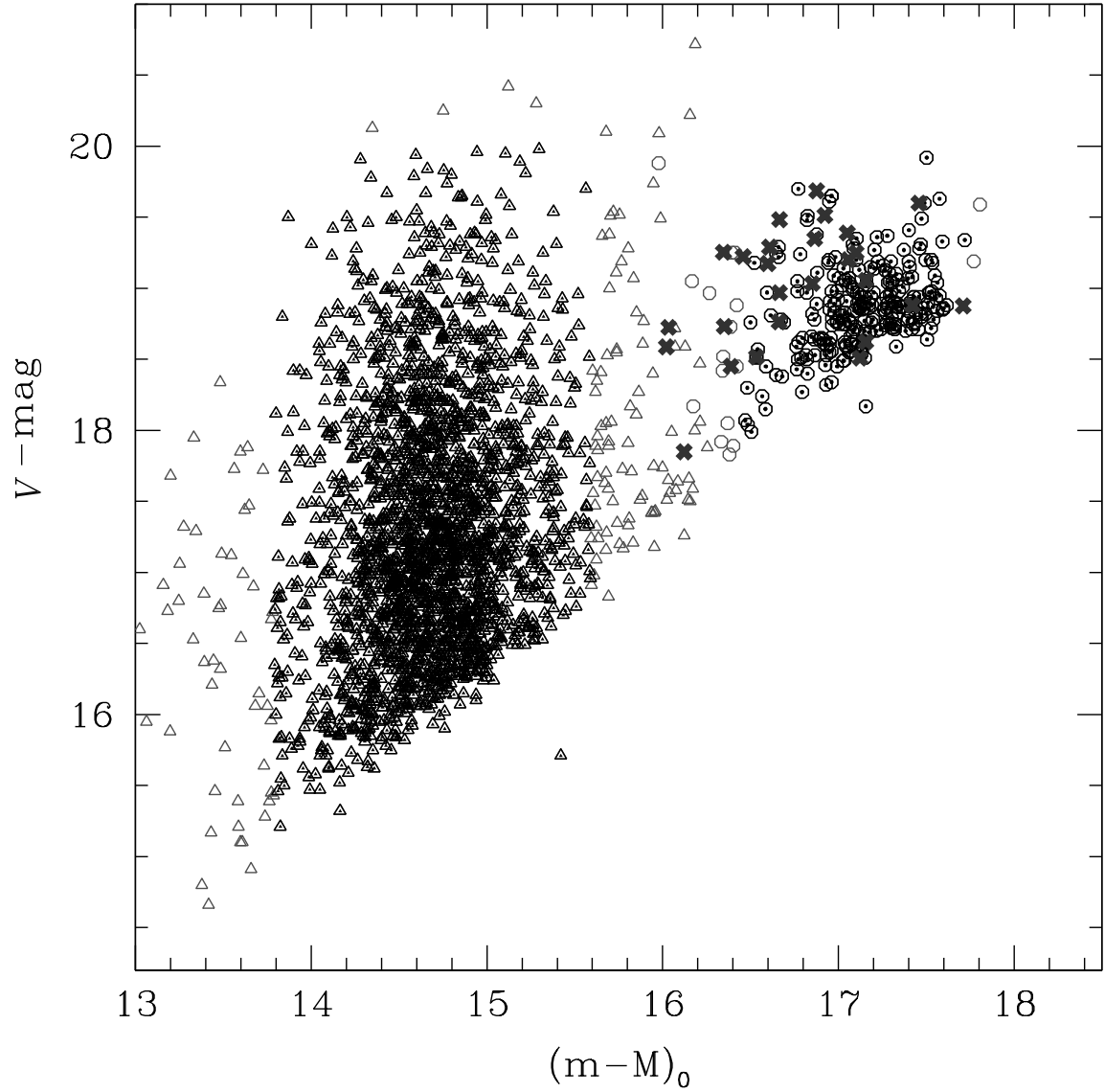}
\caption{The \citet{kunder08b} division of Bulge (triangles)
and Sgr (circles) RR0 Lyrae stars in the MACHO database.
The stars used in this analysis are indicated by symbols with
dots in the middle.
The Sgr stars used in the \citet{alc97} analysis are shown as crosses.
\label{plottwo}}
\end{figure}

The division of Bulge and Sgr RR0 Lyrae stars in the MACHO database
as determined by \citet{kunder08b} is shown in Figure~\ref{plottwo}.
Again, only the stars with photometric metallicities from \citet{kunder08a}
are plotted.  The abscissa is the distance modulus to each star, $(m-M)_0$, 
using Equation~(1) for absolute magnitude and corrected for
extinction, explained later in \S 4.
One can clearly see a concentration of stars which represent the RR Lyrae stars
located in the Bulge, and a concentration of stars which represent the Sgr 
galaxy.  However, between the
two populations there is some ambiguity as to which population
a RR Lyrae star truly belongs.  There may also be some RR0 Lyrae stars
that belong to neither the Bulge nor the Sgr galaxy, but belong
to the halo and thick disk.

\begin{figure}[htb]
\includegraphics[width=16cm]{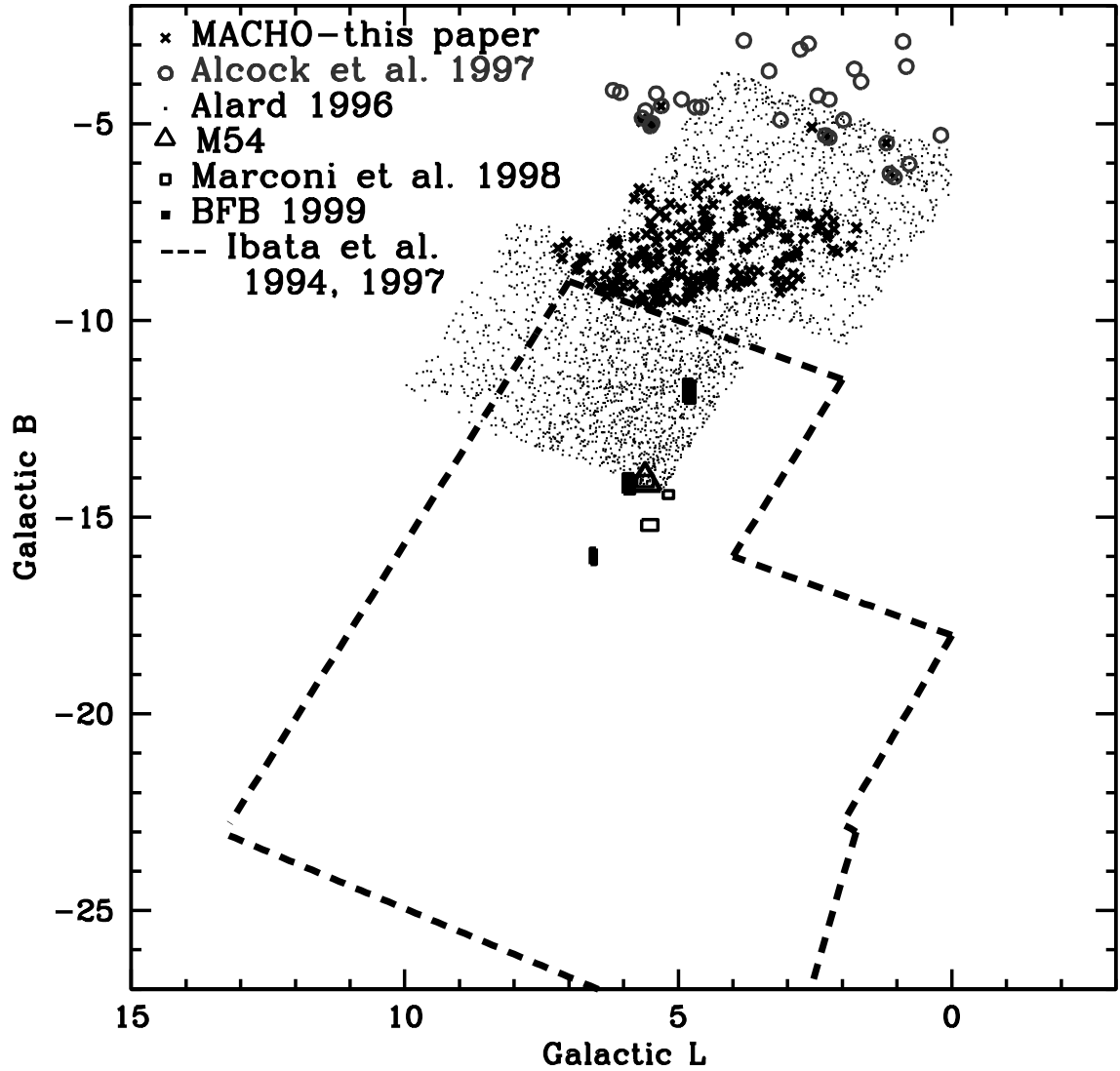}
\caption{The location of the stars used in this analysis as a function
of Galactic latitude and longitude.  Also shown are samples from
other studies with distance estimates to the Sgr galaxy, where the distances
are given in Table~\ref{DistTab}.  BFB 1999 refers to \citet{bellazzini99}.
\label{plotthree}}
\end{figure}

The relative distances between the bulge and Sgr could be dependent
on the samples used ({\it i.e.,} if brighter bulge stars are
included in the sample, the average distance to the bulge would
be smaller).  To ensure a consistent and accurate bulge and Sgr sample, the 
standard deviation of the extinction corrected distance modulus for
each population is found.  The stars that are within 2.0 $\sigma$ of the
mean of each distribution 
are indicated in Figure~\ref{plottwo} by symbols with dots in the middle.
Other cuts that encompass 1.5$\sigma$ and 1.0$\sigma$ of each distribution
and that include the stars brighter than 19.1 mags are investigated later 
in this paper.
It is evident from Figure~\ref{plottwo} that the RR0 Lyrae stars in
the \citet{alc97} sample tend to have a smaller $(m-M)_0$ than the
majority of the Sgr RR Lyrae stars used here.  These stars also have
Galactic latitude values that place them closer to the Galactic plane.
Hence it is unclear if the \citet{alc97} sample is biased to include 
Sgr stars that have on average closer distances, or if Sgr RR0 Lyrae 
stars with smaller |b| values are truly closer to us.  Figure~\ref{plotthree} 
shows the location of the \citet{alc97} sample, the MACHO RR0 Lyrae 
star sample used in this paper, and a number of other relevant samples 
from studies with distance estimates to the Sgr galaxy, as a function 
of Galactic lattitude and longitude.

The RR0 Lyrae stars are binned according to MACHO field, so the relative 
distance between the bulge and Sgr in each MACHO field can be found.
Although all MACHO bulge fields contain an ample number of RR0 Lyrae stars 
in the \citet{kunder08b} sample, 
only the MACHO fields at lower galactic latitudes ($\rm|b|<5^{\circ}$) 
contain a significant amount RR0 Lyrae stars that belong to the Sgr Galaxy.  
This analysis is restricted to MACHO fields containing three or more Sgr 
stars in order to minimize small number statistics and unknown reddenings. 

\begin{figure}[htb]
\includegraphics[width=16cm]{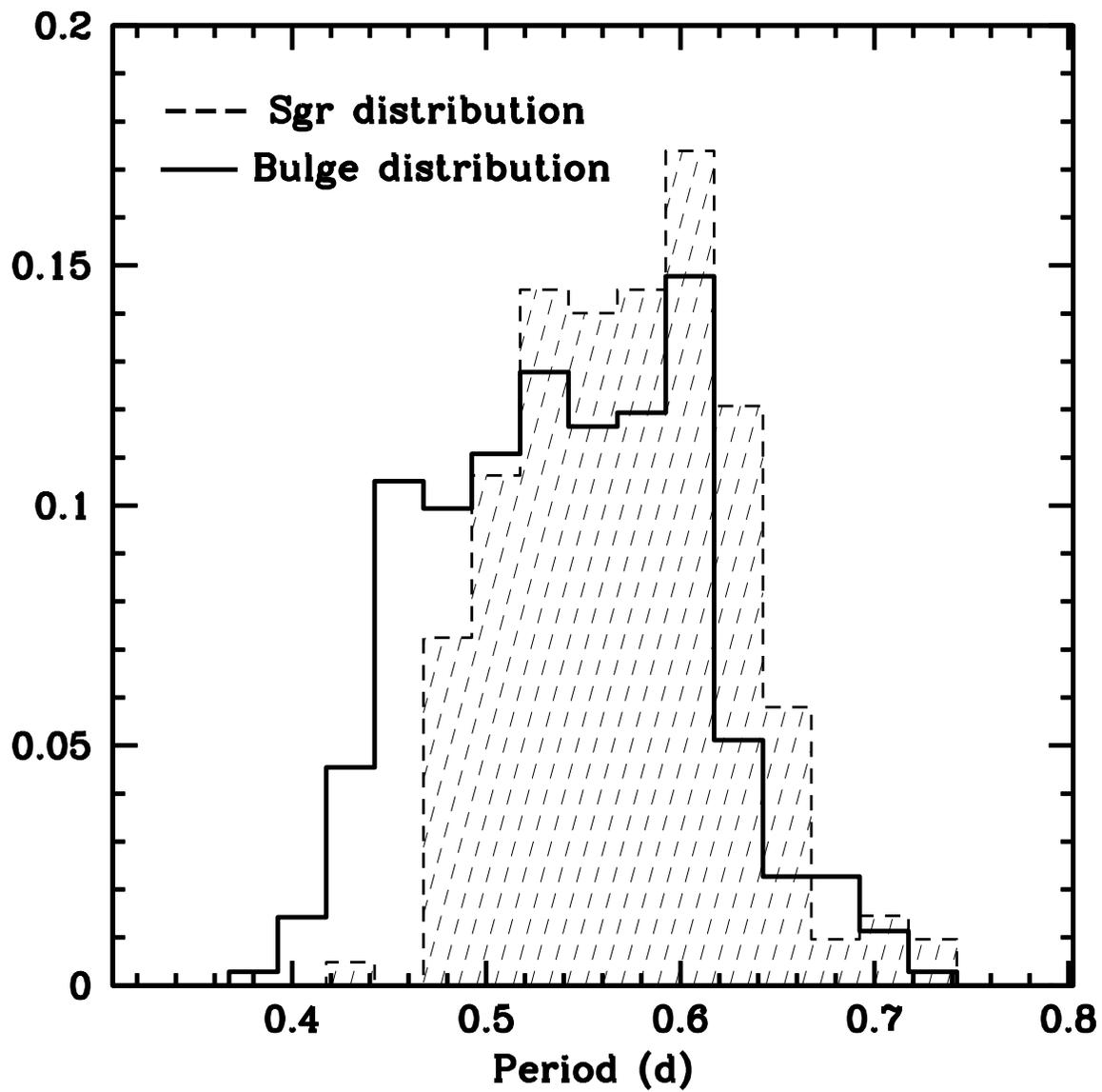}
\caption{This is a normalized histogram of 352 Galactic Bulge (solid) and
207 Sgr (dashed) RR0 Lyrae stars' periods.
\label{plotfour}}
\end{figure}

\begin{figure}[htb]
\includegraphics[width=16cm]{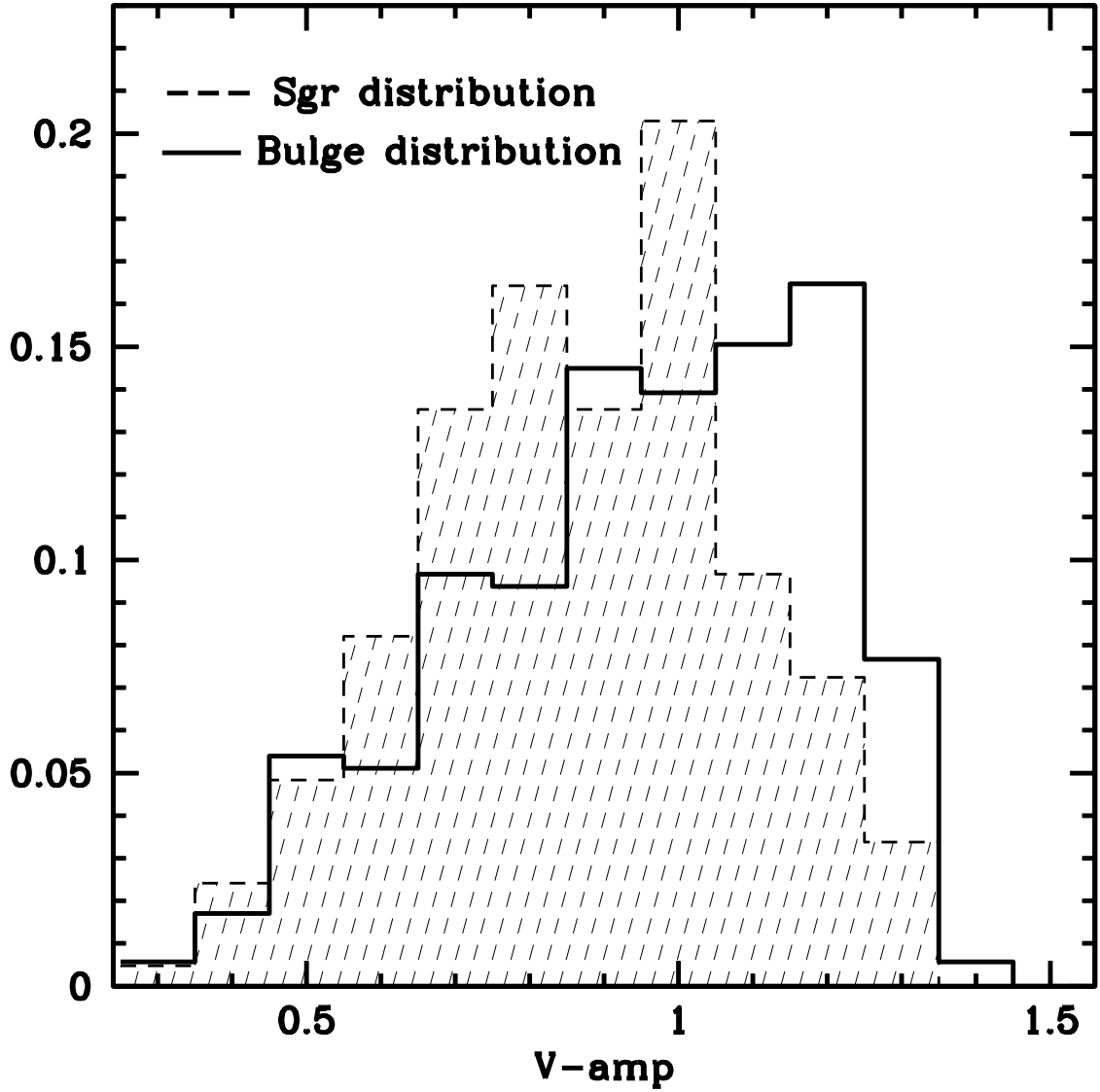}
\caption{This is a normalized histogram of 
352 Galactic Bulge (solid) and
207 Sgr (dashed) RR0 Lyrae stars' $V$-amplitudes. 
\label{plotfive}}
\end{figure}

Figure~\ref{plotfour} and Figure~\ref{plotfive} show the normalized
period and $V$-amplitude distribution of the Bulge and Sgr RR0 Lyrae stars
in MACHO fields containing 3 or more Sgr stars.  The
\citet{cseresnjes01} period analysis of $\sim$3700 RR Lyraes 
distributed between Sgr and the Milky Way found that
although the RR Lyrae stars in Sgr present the shortest average periods among
all the dwarf galaxies, their periods are still on average longer
than the RR Lyrae stars in the Galactic Center.  This is evident in 
Figure~\ref{plotfour} as well.  Because the Sgr stars are fainter, 
it would be harder to detect low amplitude stars in the Sgr sample.  
However, the $V$-amplitude distribution of the Bulge and Sgr stars
looks similar, and lends credence to the completeness of the Sgr sample. 
In order to assure that the RR0 Lyrae stars in the Bulge and the Sgr can be 
inter-compared without any potential bias,
the relative distance between the RR0 in the bulge and in Sgr is
computed here using the RR0 in the bulge covering the same
period range as the Sgr RR0 Lyrae ({\it i.e.}, 0.46d $<$ P $<$ 0.66d).
This period cut has only a minor effect on the \feh of the sample.

\begin{figure}[htb]
\includegraphics[width=16cm]{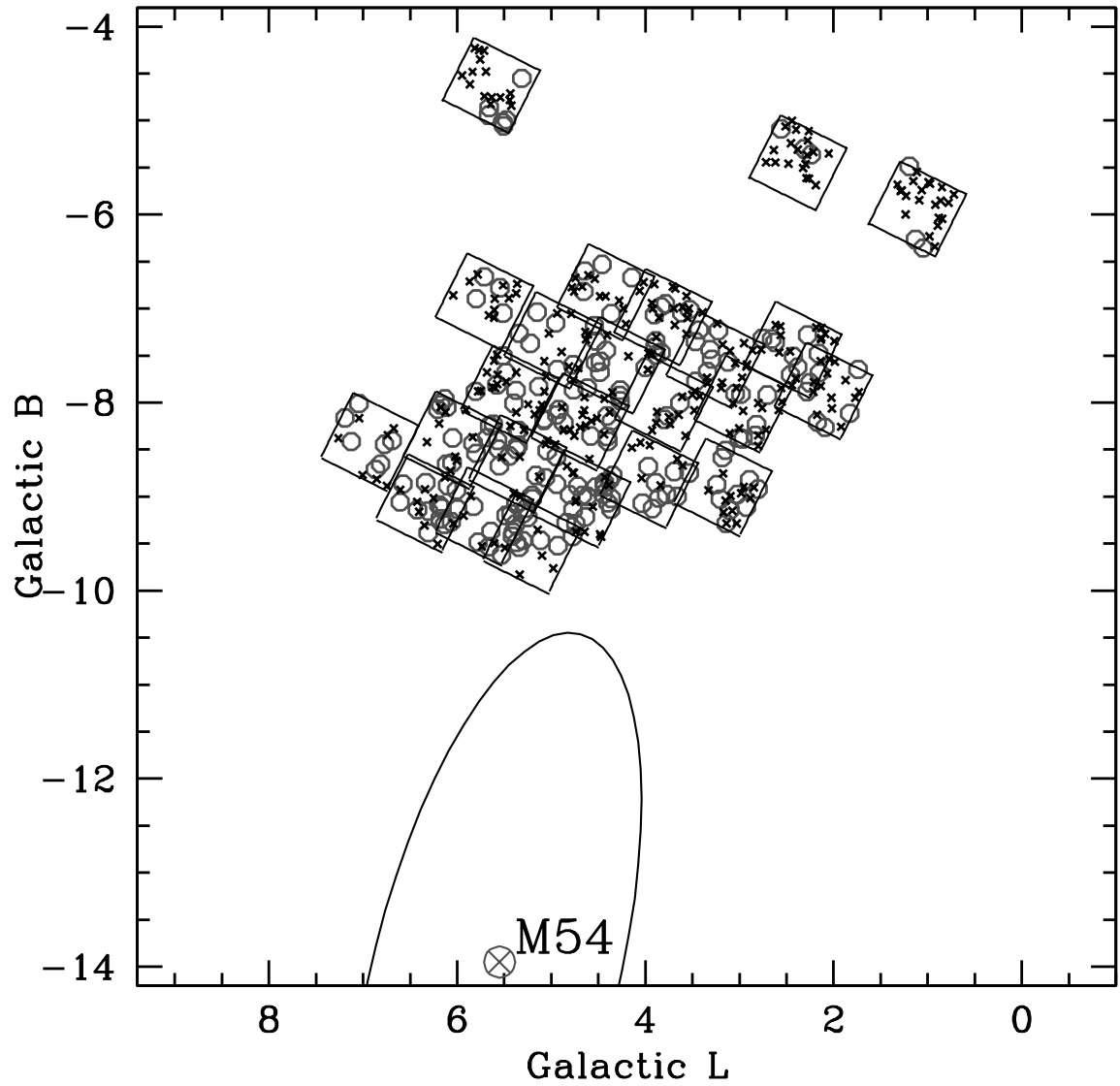}
\caption{The location of 288 Bulge RR0 Lyrae stars (crosses) and
203 Sgr RR0 Lyrae stars (circles) from the MACHO bulge fields.
Only RR0 Lyrae stars within the period range of the Sgr RR0 Lyrae
stars are shown in this figure, and only the MACHO fields containing 
three or more Sgr stars are shown and considered in this paper.  Also
shown is the location
of the globular cluster, M54, which is at the center of the Sgr galaxy,
and the location of the main body of Sgr, as traced out by M giants 
from the 2MASS survey.
\label{plotsix}}
\end{figure}

Figure~\ref{plotsix} shows
the location of the MACHO Bulge and Sgr stars in MACHO fields
with three or more Sgr stars and that have the above period range.  There 
are 288 Bulge
and 203 Sgr RR0 Lyrae stars in the MACHO fields that satisfy these
criteria.  The
position of the globular cluster, M54, located at the center of the Sgr 
galaxy is indicated as well as the core radius of the Sgr galaxy as traced
 out from M giants \citep[assuming an ellipticity of 0.65 and a position 
angle of 104$^\circ$;][]{majewski03}.

\section{Reddening}
The reddening is patchy in the MACHO fields toward the bulge, and
on large scales, extinction is regularly stratified parallel to the
Galactic plane.  
\citet{kunder08b} show that the apparent $(V-R)$ color of RR0 
Lyrae stars at minimum $V$-band light can be utilized to
measure the amount of interstellar reddening along the line of sight to the 
star since the intrinsic $(V-R)_0$ colors at minimum $V$-band light
seem constant.  They further provide evidence that the intrinsic
color at minimum light is very insensitive to metallicity and the 
Blazhko effect.  The reddening values derived from their procedure
for the Sgr and Bulge stars are used here.  The average E(V-R) for
the Bulge RR Lyrae stars is 0.24 $\pm$ 0.04 and the average E(V-R) for the
Sgr sample is 0.26 $\pm$ 0.04.  Using the selective extinction coefficient
$R_{V,VR}=A_V/E(V-R) = 4.3$ \citep{kunder08b}, the average $V$-band 
extinction is one magnitude.
 
\begin{figure}[htb]
\includegraphics[width=16cm]{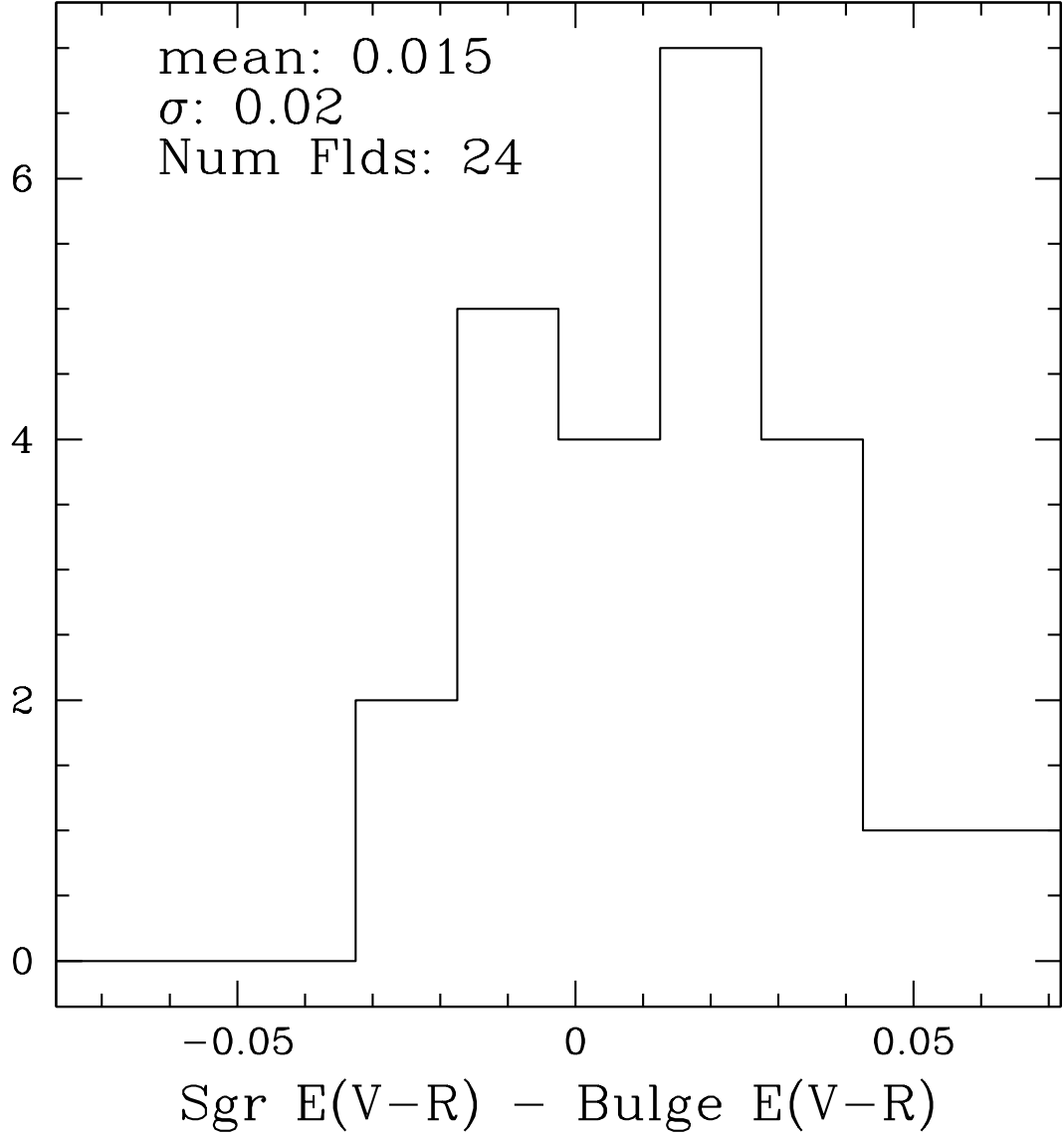}
\caption{Histogram of the difference in the average MACHO field Bulge 
and Sgr color excess, $\rm Bulge_{E(V-R)} - Sgr_{E(V-R)}$.
\label{plotseven}}
\end{figure}

In order to adopt an accurate reddening estimate,
first a check on how the reddening differs from RR0 Lyrae stars
in the Bulge and the Sgr Galaxy is performed.  
The color excess, E(V-R), along the line of sight
to each RR0 Lyrae star is calculated using its (V-R) color at 
minimum $V$-band light.  The E(V-R) values of the Galactic Bulge 
and Sgr RR0 Lyrae stars in each MACHO field are averaged
together and the difference in the Bulge and Sgr color excess is 
shown in Figure~\ref{plotseven}.  It is suggestive that 75\% of the
E(V-R) values are positive, which means that the stars of the Sgr
are on average slightly more reddened than the stars in the Bulge.
The negative values on the plot are unphysical, as that would mean 
the Sgr stars are closer to us than the Bulge.  From these
negative values, we take the uncertainty in the color excess 
within each field to be $\sim$ 0.015 mags.

The difference in the Bulge and Sgr color excess as a function of
Galactic Latitude and as a function of Galactic Longitude 
were examined.  No trend was found.
To determine the extinction in the $V$-band, the
color excess along the line of sight of the Bulge and Sgr RR0 Lyrae stars 
in each MACHO field is averaged 
and multiplied by the selective extinction coefficient.

\section{Distance as a Function of Position from ($l,b$) = (0$^{\circ}$,0$^{\circ}$)}
\subsection{A Triaxial Bulge}
\begin{figure}[htb]
\includegraphics[width=16cm]{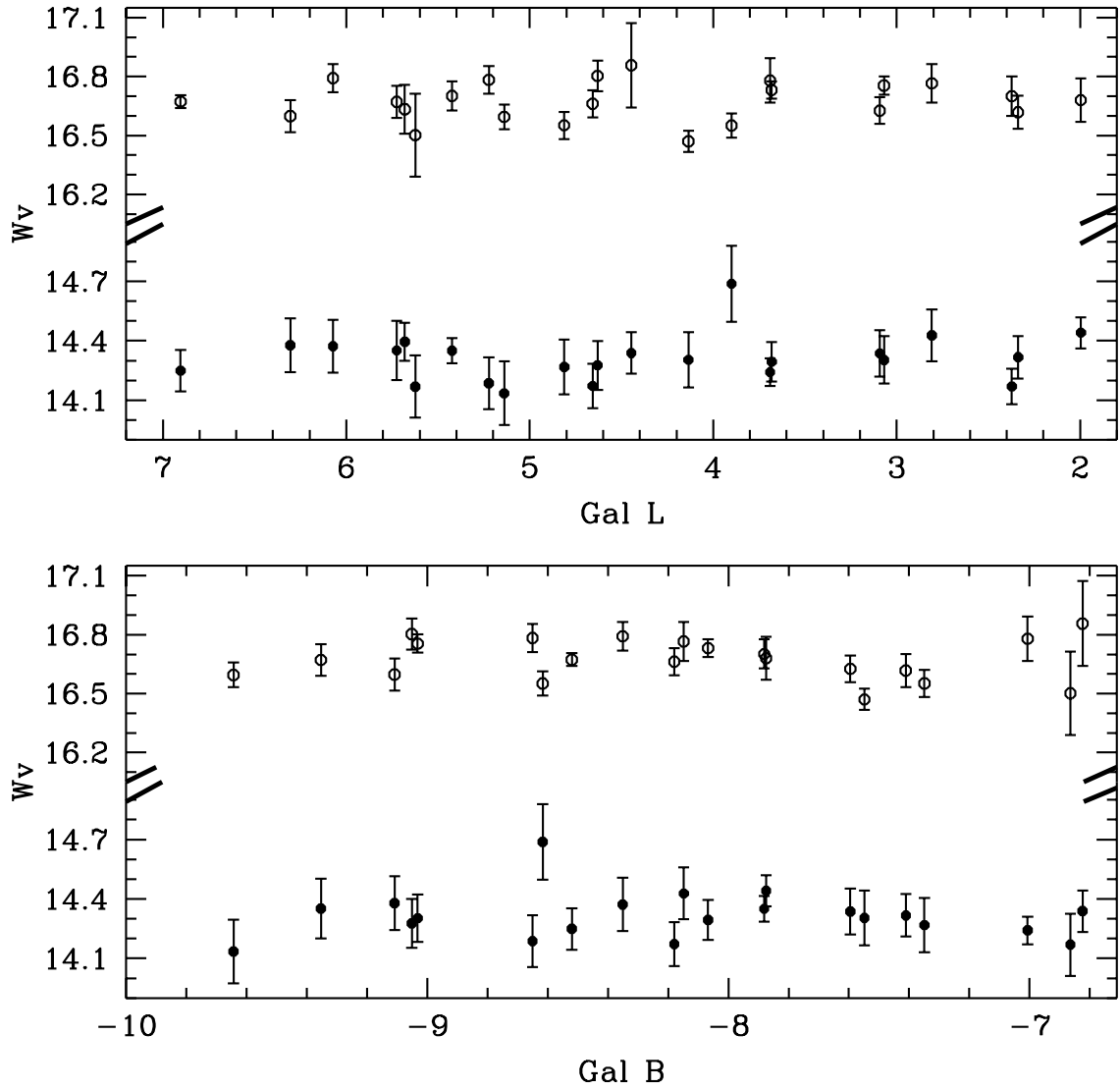}
\caption{Mean reddening-independent magnitudes of RR0 Lyrae stars
used in this analysis versus Galacitic $l$ and $b$.  The RR0 Lyrae stars 
are binned by MACHO field.  The mean bulge $W_V$ magnitudes are 
represented with filled circles, while the Sgr RR0 Lyrae $W_V$ magniutes 
are shown as open circles.  There is no trend with Galactic $l$
in the Bulge RR0 Lyrae star sample, such as found with the Bulge red clump 
giants\citep{alc98, stan94, cabrera07}.  Note the break in the Bulge and Sgr $W_V$ range, between 14.9 and 16.1
mags.
\label{ploteight}}
\end{figure}

It is well known that the bulge of the Milky Way is triaxial \citep[e.g.,]
[and references therein]{lopez05,picaud04}.  For a barred distribution 
with a standard inclination angle, stars at a larger longitudes
would be nearer and hence brighter, than those at smaller longitudes.  
The MACHO Bulge RR0 Lyrae stars span a range of Galactic $l$ and $b$, and
as the distance to Sgr is determined in a differential way, comparing
the magnitude of RR Lyrae stars in Sgr and in the Bulge,
the effect of a triaxial bulge on the MACHO RR0 Lyrae stars is
investigated.  Figure~\ref{ploteight} shows the mean reddening-independent 
magnitudes in each MACHO field for the stars used in
this analysis.  Reddening-independent magnitudes are defined as 
$W_V = V - 4.3(V-R)$, where the factor 4.3 is the selective extinction coefficient 
$R_{V,VR}$ derived by \citet{kunder08b}.  The errorbar is the dispersion in the 
mean $\rm W_V$ of the stars in each field.  There is no trend in $\rm <W_V>$ as a
function position, which is what would be expected if the RR0 Lyrae stars
traced out the barred distribution in the Bulge.  This is not surprising;
\citet{kunder08a} find no strong bar signature when restricting the 
MACHO RR0 Lyrae sample to those stars closest to the Galactic plane.
\citet{collinge06} find a weak barred signature in the OGLE Bulge 
RR0 Lyrae population and \citet{alard96}, \citet{alc98}, and 
\citet{wesselink87} also find no strong bar in the RR Lyrae distribution.  
It is generally assumed that the absence of a strong bar in the bulge RR Lyrae 
suggests that these stars represent a different population than 
the majority of the more metal-rich stars in the bulge.  

\subsection{A Model Bulge}
Translating the heliocentric distance of a star to the galactic center, $R_0$,
involves sin $b$ for $l$=0$^{\circ}$, and more complex relations for $l\ne0^{\circ}$.
The MACHO fields are not located directly behind the center of the bulge 
but at a Galactic latitudes as low as $-$10$^{\circ}$, and all the MACHO fields in
this analysis have $l<$0$^{\circ}$,.  In order to determine
how substantial an effect this is, we adapt the procedure used by
\citet{carney95}, who modeled the expected RR Lyrae density versus 
distance in Baade's window using:  
\begin{equation}
dN = dR\ A_{eff} N_0\ \rm{cos}\ b[X^2 + Y^2 + (Z/k)^2]^{\lambda/2},
\end{equation}
where $R$ = distance from observer along line of sight; $R_0$=distance to 
galactic center; $N_0$=constant (kpc$^{-3}$); $A_{eff}$=effective angular size of
each field, $\lambda$=power law exponent ($\rm<$0) of the number density; 
$X$=$R_0$-$R$cos $b$ cos $l$; $Y$ = $R$ cos $b$ sin $l$; 
$Z$ = $R$ sin $b$; and $k$=the ellipticity parameter, the ratio of the bulge
minor and major axes.  

\begin{figure}[htb]
\includegraphics[width=16cm]{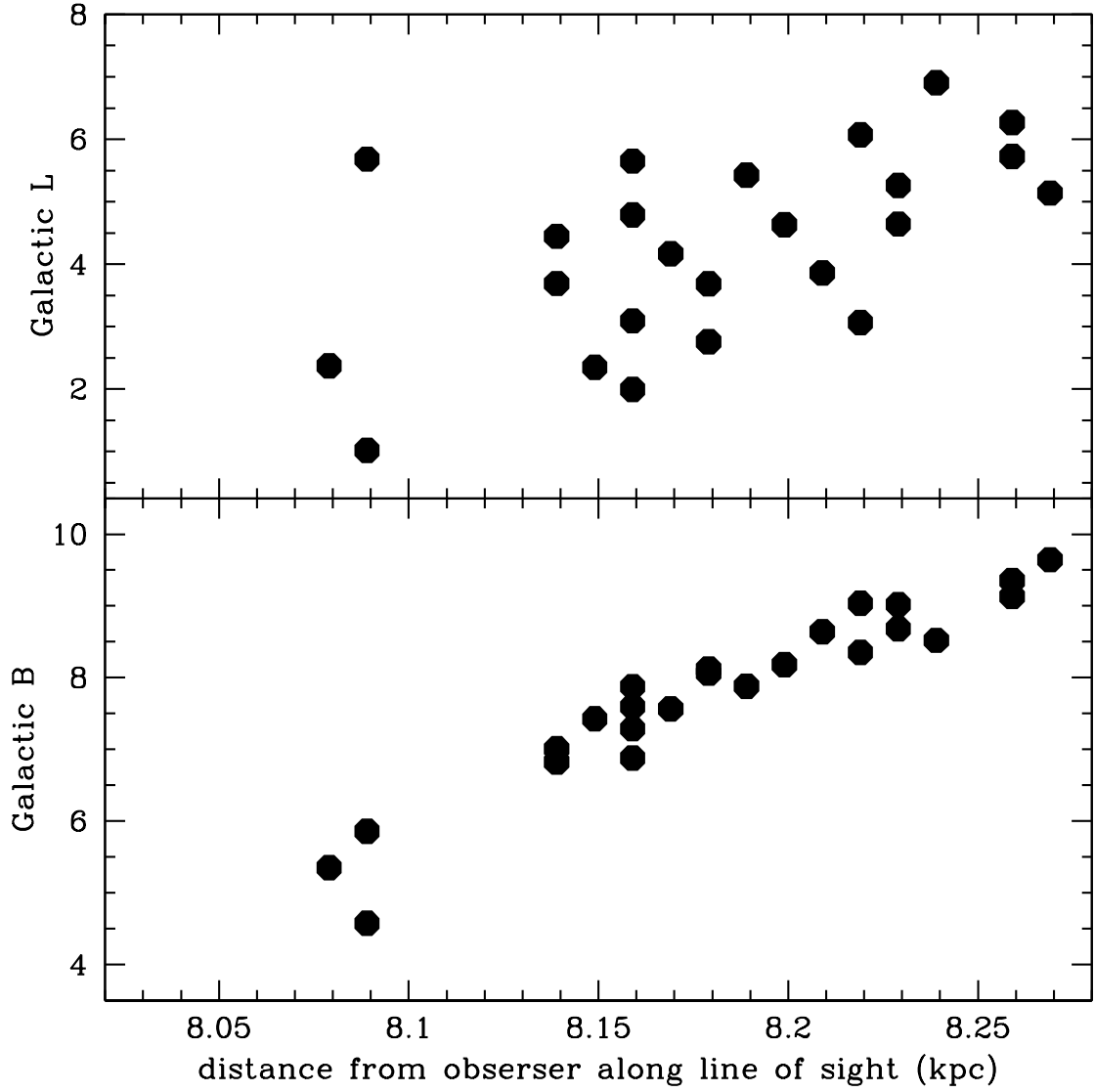}
\caption{The distance along the line of sight as a function of the $l$ and 
$b$ of the MACHO fields used in this paper.
\label{plotnine}}
\end{figure}

For each field with a unique $l$ and $b$, we assume $R_0$ = 8 kpc and vary 
$R$.  The $R$ at maximum density is the distance along the line
of sight at ($l,b$)=(0$^{\circ}$,0$^{\circ}$) (for $R_0$ = 8 kpc).
Figure~\ref{plotnine} shows how the distance from the observer along the 
line of sight varies as a function of the Galactic $l$ and $b$ values of the 
MACHO fields.  A $\lambda$=$-$2.0 is used, which is the value \citet{carney95} 
finds best fits the RR0 Lyrae data in Baade's Window, $(l, b) = 
(1 \hbox{$.\!\!^\circ$} 0, -3 \hbox{$.\!\!^\circ$} 9)$.  A $\lambda$=$-$2.3, 
which is also found by \citet{carney95} to yield satisfactory results, does 
not change Figure~\ref{plotnine} much.  A $k$=0.8 is used, which suggests a 
moderately flattened bulge.  This is the value \citet{carney95} finds yields 
"superior results" in all cases to the RR0 Lyrae data.  Although the $COBE$ 
Diffuse Infrared Background Experiment found $k \sim$0.6 
in their observations of the Galactic Bulge\citep{weiland94}, they also find
asymmetries in bulge brightness which are consistent with a triaxial bar located
at the center of the Galaxy.  $COBE$ probed $all$ stars in the Galactic Bulge 
and did not differentiate between the old, metal-poor populations, such as 
the RR0 Lyrae stars in which at best only a slight bar signature is seen, 
and the younger, metal-rich populations which are more common and more 
luminous in the bulge.  A change in $k$ from $k$=0.8 to $k$=0.6 changes the 
distance from the observer along the line of sight by +0.15 to 0.25 kpc.  
From the previous section in which no bar was seen in the RR Lyrae sample, 
it is unlikely that $k$=0.6 for the RR Lyrae population in the bulge.

The correction in the distance due to the fact that the MACHO fields are 
not at ($l, b$) = (0$^\circ$, 0$^\circ$) is a relatively small effect 
($\sim$ 0.2 kpc).  
We take this into account when using the reference 
distance to the Bulge for each MACHO field, as given in Figure~\ref{plotnine}. 

\section{Distance determination}
The difference in the average distance modulus of each MACHO Bulge and Sgr
field is found:
\begin{eqnarray}
\nonumber
\Delta (m-M)_0 &=& (<V_{Bul,RR}> - <V_{Sgr,RR}> )\\
\nonumber
&+& (<M_{V_{Sgr,RR}}> - <M_{V_{Bul,RR}}>)\\
&+& (<A_{V_{Sgr,RR}}> - <A_{V_{Bul,RR}}>).
\label{distance}
\end{eqnarray}
In the above equation, $<V_{Bul,RR}>$ and $<V_{Sgr,RR}>$ is the average 
MACHO mean $V$-band magnitude of the stars in each Bulge and Sgr MACHO field, 
respectively.
$<M_{V_{Sgr,RR}}>$ and $<M_{V_{Bul,RR}}>$ is the average absolute magnitude 
of the Sgr and Bulge stars in each MACHO field, respectively,
determined using the stars' metallicity and Equation~(1).  
$<A_{V_{Sgr,RR}}>$ and $<A_{V_{Bul,RR}}>$ is the average $A_V$ of the 
RR0 stars in each MACHO field, determined from the 
RR0 Lyrae's color at minimum light as described in the previous section.
The error in the derived distance modulus included the error in the 
photometry, the uncertainty in the ratio of selective to total extinction,
and the error in the reddening for both the Sgr and Bulge stars.  The
reliability of this error estimate was confirmed by using the
small sample statistical formulae of Keeping (1962, p. 202)
to calculate the standard error of the mean of the distance modulus
in each MACHO field of the Sgr and Bulge stars.





\begin{figure}[htb]
\includegraphics[width=16cm]{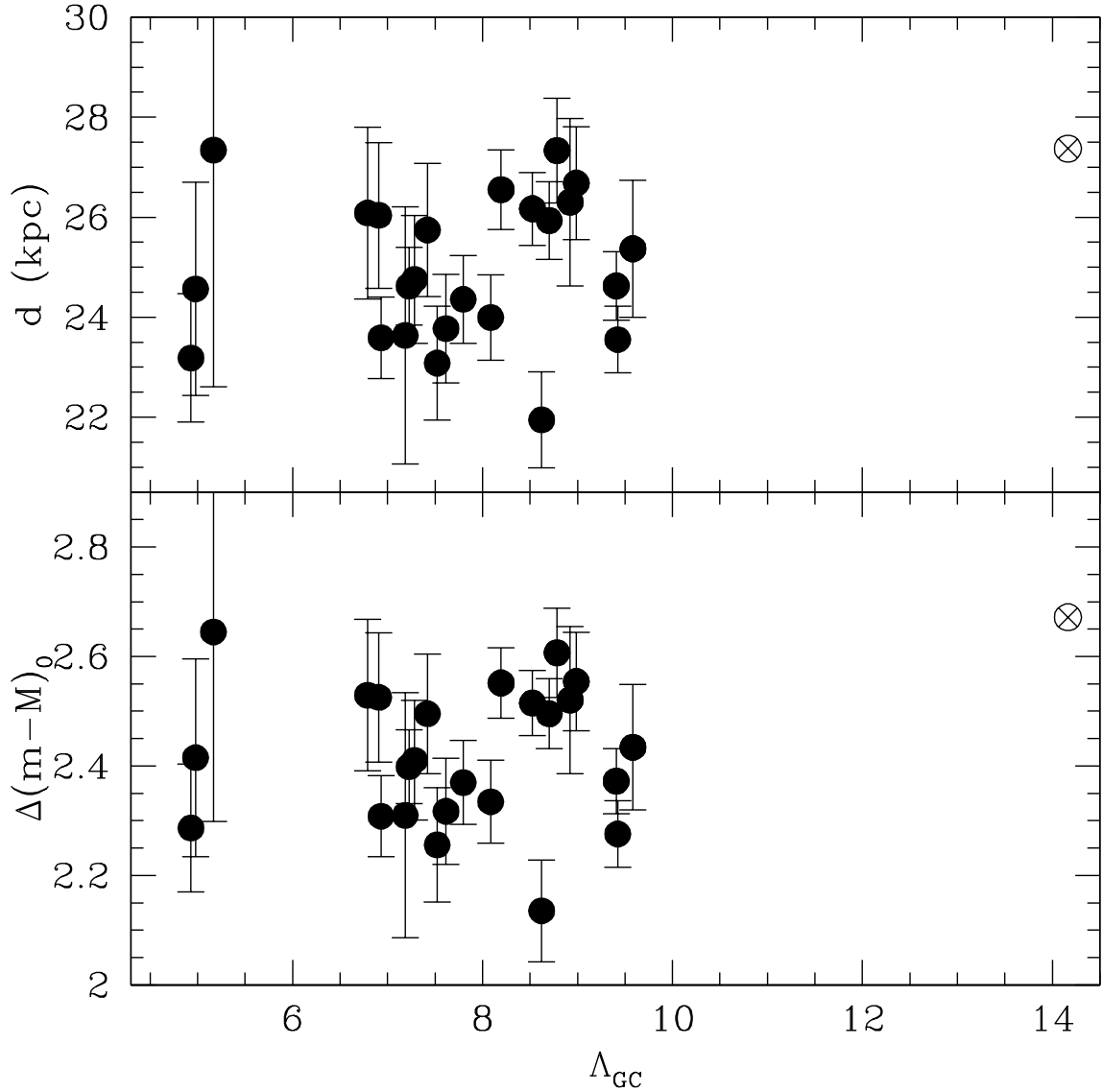}
\caption{{\textit {Bottom:}}  The difference in the distance modulus of the 
Bulge and Sgr RR0 Lyrae stars is binned according to MACHO field and shown here
as a function of $\Lambda_{GC}$, an angle in the Galactocentric spherical coordinate Sgr system
($\Lambda_{GC}$=0 at the Galactic plane).
{\textit {Top:}}  Same as below, but here the difference in the Bulge and 
Sgr RR0 Lyrae distance modulus is translated into a distance by
assuming $\rm R_{GC}$ = 8 kpc.  The circle at $\Lambda_{GC} \sim 14^{\circ}$
represents M54.
\label{plotten}}
\end{figure}

The differences of each MACHO fields' distance modulus of the bulge
and Sgr RR0 Lyrae stars are shown in Figure~\ref{plotten} as a function
of $\rm \Lambda_{GC}$, an angle in the Galactocentric spherical coordinate 
system\footnote{The standard Galactic 
coordinate system is converted to the Sgr longitudinal coordinate system 
using the C++ code from \citet{law05} }.  This is a more natural spherical 
coordinate
system for the interpretation of Sgr tidal debris, using the Sgr orbital plane
traced out by the 2MASS M giant population from \citet{majewski03}.  There are
24 data points in this figure, since there are 24 MACHO fields with
3 or more Sgr RR0 Lyrae stars.  The distance to M54, the globular cluster
located at the center of Sgr, is found
using the photometry of RR0 Lyraes from \citet{layden00}.  The reddening
was determined from $(V-I)$ color at minimum light, just as the reddening
in this analysis uses the $(V-R)$ colors at minimum $V$-band light
The absolute magnitude of these stars was determined using Equation~(1)
in an identical manner as in this analysis.  This places the distance to
M54 approximately on the same scale as the Sgr RR0 Lyraes in this paper.

We experimented with different divisions of the MACHO Sgr and Bulge
populations, particularly cuts that are within 1.0$\sigma$ and 
1.5$\sigma$ of the mean of each 
distribution, cuts that include the stars brighter than 19.1 mags, and
cuts that encompass the full period range of the Bulge RR0 Lyrae stars.
Table~\ref{tab1} summarizes these results.  It is striking that the 
various cuts do not affect the derived distance (with a range of
$\rm \Delta{\mu}$ = 2.39 to 2.42), indicating
that the method does not introduce important biases or selection effects
to the sample.

The average difference in the Bulge and Sgr distance in Table~\ref{tab1} is 
$\rm \Delta(\mu_{Sgr} - \mu_{Bulge})$ $=$ 2.41 mags with a dispersion
of 0.14 mags.  
Setting the distance to the bulge as 8 kpc 
(Groenewegen, Udalski \& Bono 2008, Eisenhauer et al.\ 2005), we find 
the distance to the Sgr galaxy is 24.8 kpc $\pm$ 0.8 kpc (internal).
This difference D = 24.8 kpc is significantly different from the distances
of the 63 M54 RR0 Lyrae stars measured from \citet{layden00}.
If this distance spread between the RR Lyrae in M54 and the RR
Lyrae in the MACHO fields (located at  approximately $l$ = 5$^{\circ}$ and 
$b$ = -8$^{\circ}$) is real, it would mean that the Sgr is inclined along
the line of sight.   

This estimate is quite a bit larger ($\sim$2.0 kpc) than that from 
\citet{alc97}, who uses MACHO RR0 Lyrae stars and an approach similar to 
that performed here.  However, their $\sim$ 24 Sgr star sample
is located closer to the galactic plane than the sample used
here, does not correct for the line of sight of the MACHO
fields, and does not take into account the metallicity difference
between the two populations.  All of these factors have
the effect of decreasing the distance to Sgr.  

\citet{alard96} used 1466 RR0 Lyrae stars discovered in a 25 deg$^2$ field,
centered at the Galactic coordinates b = -7$^{\circ}$, l = 3$^{\circ}$, to
derive the distance to Sgr as 24 $\pm$ 2 kpc.  The location of this 
field is similar to the location of the MACHO fields, and the distance
determination is in very good agreement with that found in this paper.  Other
distance estimates are listed in Table~\ref{DistTab}; direct
comparisons are difficult to make since many of the studies differ
in significant ways, \textit{i.e.}, Sgr population, position in the sky.  
It would be
interesting to do similar differential studies using RR Lyrae stars 
that populate other locations in the Sgr galaxy.

\section{Comparison with Recent Sgr Models}
Models of the disruption of Sgr based on numerical simulations of the 
Sagittarius plus the Milky Way are available in
the literature.  Detailed comparisons are made here between the distances 
of the MACHO fields based on the RR Lyrae stars and the most recent
theoretical models: \citet{delgado04} and Law, Johnston \& Majewski (2005).

\begin{figure}[htb]
\includegraphics[width=16cm]{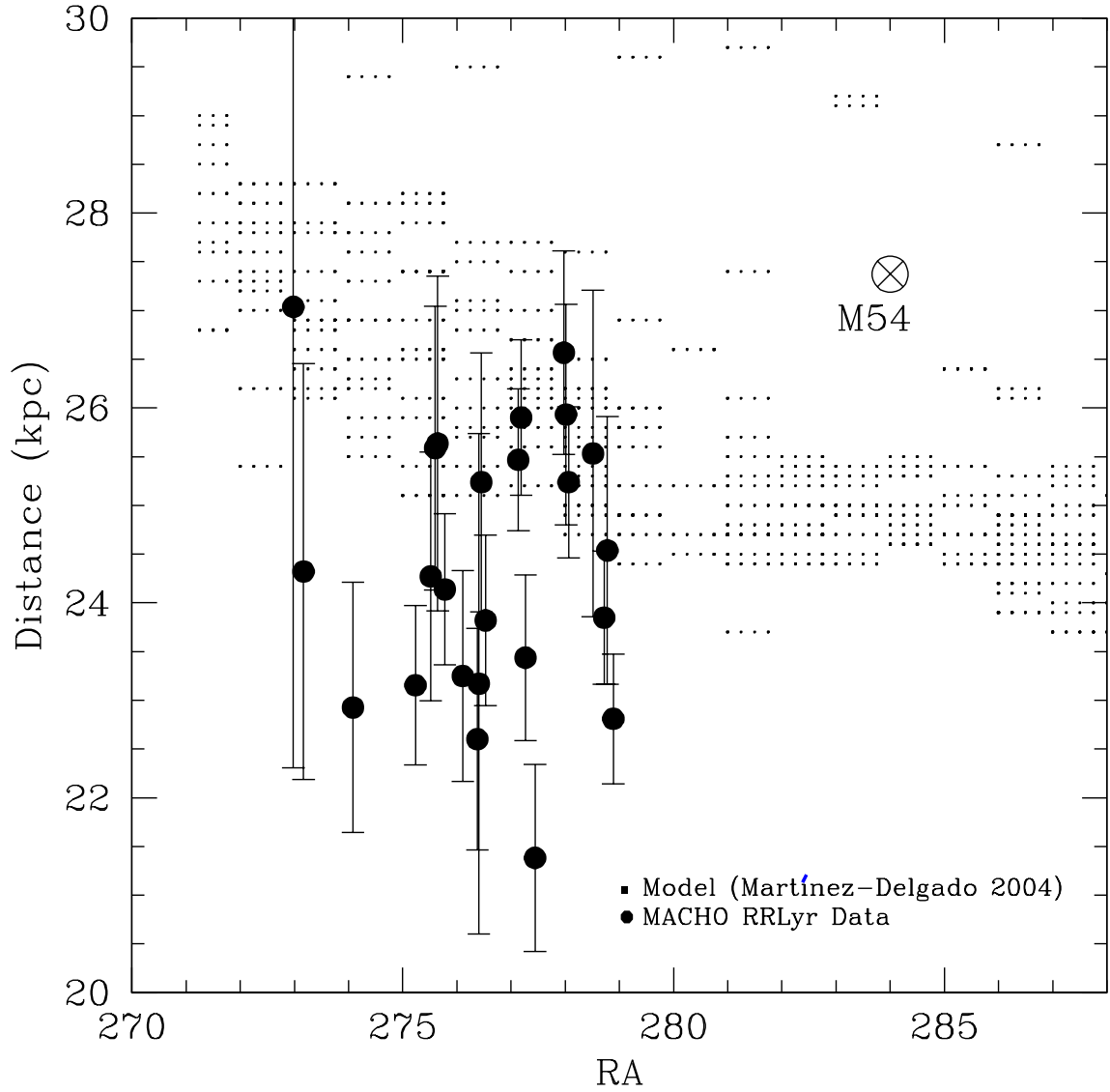}
\caption{Heliocentric distances vs. RA of the MACHO RR Lyrae
data together with the \citet{delgado04} Sgr model. 
\label{ploteleven}}
\end{figure}

Figure~\ref{ploteleven} is a plot of RA against distance for the RR Lyraes
in the MACHO survey.  The model of \citet{delgado04} (their Figure~6)
fails to reproduce in detail the location of the MACHO RR Lyre stars.  
\citet{delgado04} assumes a distance of 25 kpc for M54, whereas the 
distance to M54 determined from RR Lyrae stars is 27.3 kpc.  Although shifting 
the distance of the MACHO RR Lyrae stars by a distance of -2.3 kpc places M54
in agreement with the \citet{delgado04} model, the MACHO observations with
their slightly smaller values of RA than M54 do not overlap at all.
Vivas, Zinn, \& Gallart (2005) find a similar result with QUEST RR 
Lyrae stars, in that
the \citet{delgado04} model does not reproduce the spread of RR Lyrae
distances in the particular right ascension of the QUEST survey 
(RA $\sim$ 200-230$^{\circ}$).

\begin{figure}[htb]
\includegraphics[width=16cm]{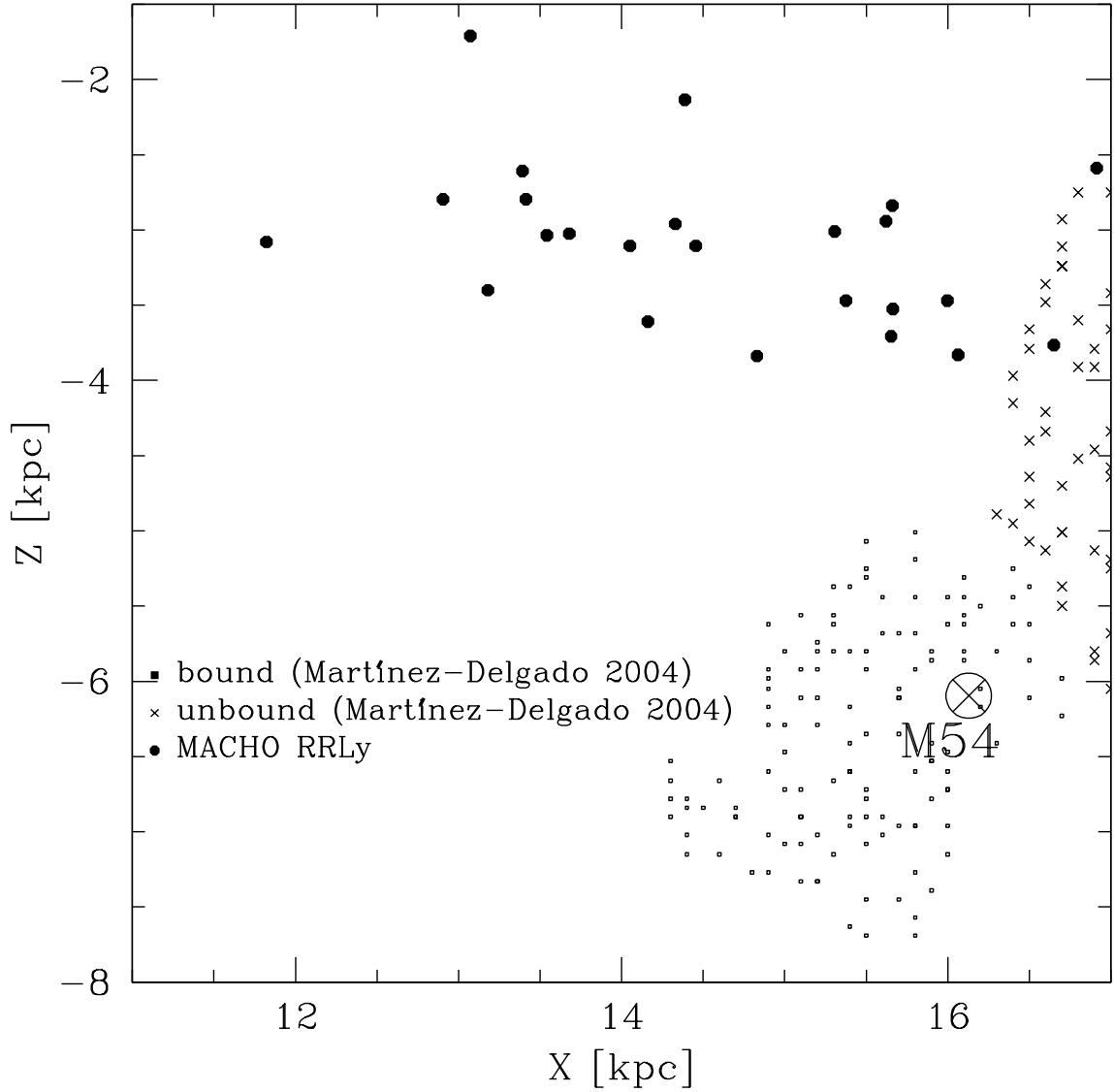}
\caption{$X,Z$ projection of the MACHO RR Lyrae data with respect to the 
Galactic center.  The Sun's coordinates are $(X,Y,Z)_\odot$=(-8.5,0.0,0.0)kpc,
and Sgr center is placed at $(X,Y,Z)_{Sgr}$=(16,2,-6)kpc.  The squares
represent particles from the \citet{delgado04} Sgr model that are still
bound to the Sgr galaxy and the crosses represent particles that 
became unbound during the last gigayear.
\label{plottwelve}}
\end{figure}

Figure~\ref{plottwelve} shows the $X,Z$ projection of the Sgr stream with 
respect to the Galactic center.  Here we have adjusted the zero-point of 
the distance modulus so that M54 corresponds to the same
approximate location as \citet{delgado04}; consequently, the MACHO
RR Lyrae data is also now placed on the same \citet{delgado04} scale.
Again the model fails to reproduce the MACHO RR Lyrae 
observational data.  For the average distance of the Sgr orbit, their 
potential flatness was an oblate halo with $q$$\sim$0.85.

\citet{law05} use M giants found in the 2MASS survey to model the Sgr
galaxy.  Figure~\ref{plotthirteen} shows the MACHO RR Lyrae observations
in the $X_{Sgr,GC},Y_{Sgr,GC}$ plane (see Majewski et~al.\ 2003 for details 
of the Cartesian Sgr,GC plane), along with the \citet{law05}
$N$-body tidal debris in the Sgr,GC plane for a spherical ($q$=1) model of the
Galactic halo potential.  Again the zero-point of the distance modulus is 
adjusted so that M54 corresponds to the distance used by \citet{law05}.  
This time the agreement between the model and the observations agrees
nicely.

\citet{vivas04} find that models that assume spherical and prolate dark
matter halos provide better fits to the QUEST data.  This appears to 
be the case for the MACHO data as well.

\section{Conclusion}  
A differential approach and RR0 Lyrae stars from the MACHO 
database are used to provide a 
new estimate of the distance modulus to the Sgr galaxy.  We take
advantage of the fact that the MACHO bulge fields have RR0 Lyrae stars
located both in the bulge and the Sgr dwarf galaxy, which can be
separated by examining their $V$ magnitudes.  By finding the relative
distances between the bulge and Sgr in each given MACHO field,
systematic effects are largely avoided.  The obtained distance
modulus is $\rm \Delta(\mu_{Sgr} - \mu_{Bulge})$ $=$ 2.41 at 
$l$ = 5$^{\circ}$ and $b$ = -8$^{\circ}$, which 
corresponds to $(m-M)_0$ = 16.97 or
D = 24.8 $\pm$ 0.8 kpc, for $\rm R_{GC}$ = 8 kpc.
This distance is significantly smaller than the distance derived from
the RR Lyrae stars located in M54 from \citet{layden00}.  This indicates
that at distances further from the body of Sgr, the Sgr galaxy
is closer to us.  Hence, the extension of the Sgr galaxy
towards the galactic plane is inclined toward us.

\begin{figure}[htb]
\includegraphics[width=16cm]{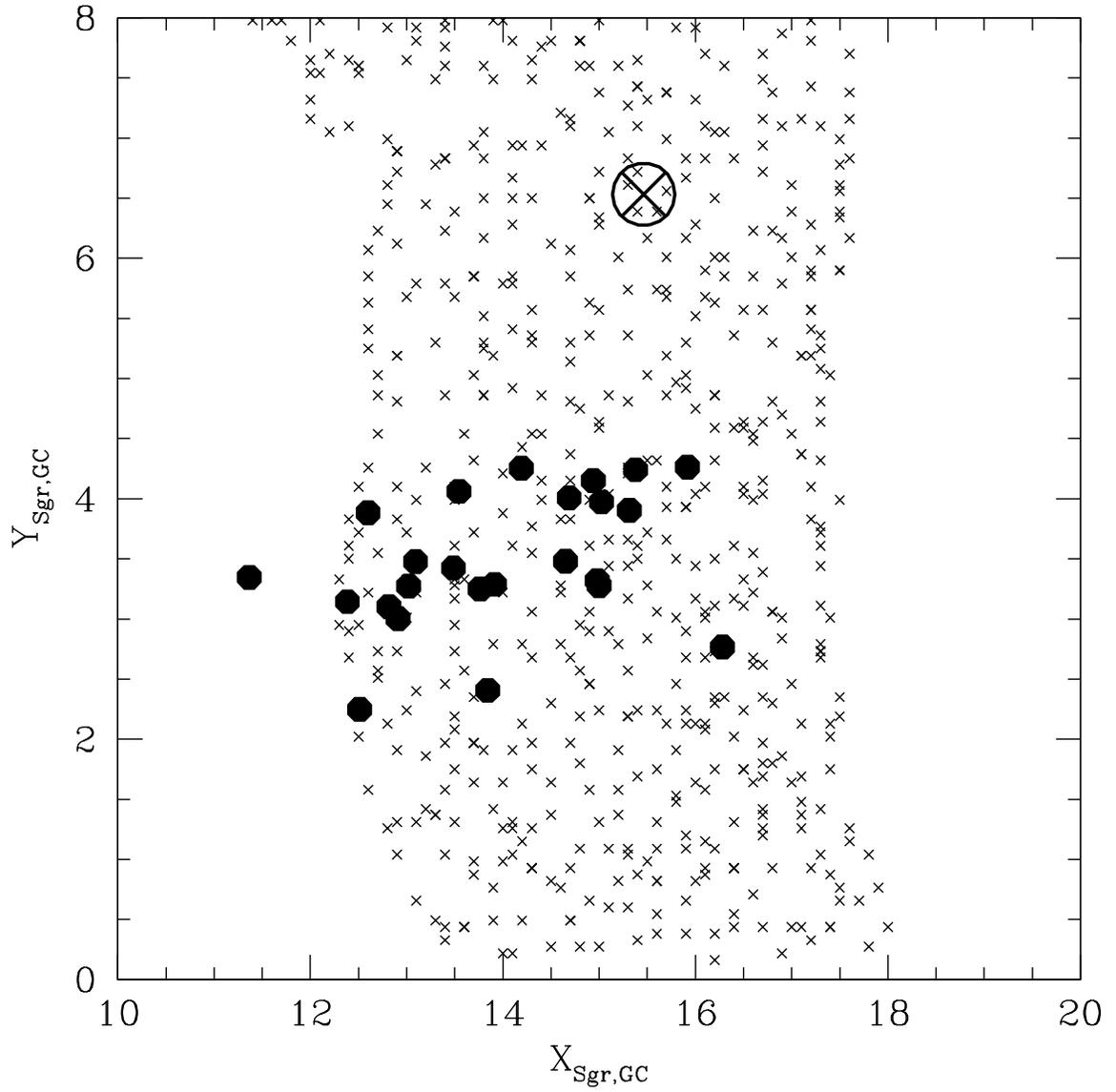}
\caption{MACHO RR Lyrae data on the Sgr,GC plane (\textit{filled circles})
along with the $N$-body tital debris model (corresponding to $q$=1.0
model) discussed by \citet{law05}.  The location of M54 is indicated 
by a circle with a cross in the middle
\label{plotthirteen}}
\end{figure}

Differential studies have the advantage of canceling many
systematic effects that occur in data collection, reduction
and analysis. 
Given the small error bar in the distance estimate determined here for 
the Sgr Galaxy, models that trace out
the orbit of Sgr and determine its previous history can be more
tightly constrained.  Our observations are compared to recent models
of the destruction of the Sgr galaxy.  Models that assume 
an oblate flattening of the dark matter halo provide a poor fit to the
data \citep{vivas04}.  Models that assume spherical dark matter halos 
($q$ $=$ 1.0, as shown in Figure~\ref{plotthirteen}) agree better
with the MACHO RR Lyrae observations.
\acknowledgments

We thank the referee for insightful comments that helped us improve the paper
and strengthen the presentation of our results.

\clearpage
\begin{table}
\centering
\caption{Average relative distance between the bulge and Sgr RR0 Lyrae stars.}
\label{tab1}
\begin{tabular}{lccccc} \\ \hline
Cut - All stars & $\rm \Delta (m-M)o_{All Stars}$ & Std Deviation &  $\rm \Delta (m-M)o_{Period Cut}$ & Std. Deviation \\ \hline
1 $\sigma$  & 2.41 & 0.11 & 2.42 & 0.10\\ 
1.5 $\sigma$ & 2.42 & 0.13 & 2.42 & 0.12\\
2 $\sigma$ & 2.42 & 0.14 & 2.42 & 0.12\\
Cut - $V$-mag $<$ 19.1 & $\rm \Delta (m-M)o_{All Stars}$ & Std Deviation &  $\rm \Delta (m-M)o_{Period Cut}$ & Std. Deviation\\ \hline
1 $\sigma$ & 2.40 & 0.12 & 2.41 & 0.12\\ 
1.5 $\sigma$ & 2.41 & 0.15 & 2.42 & 0.14\\
2 $\sigma$ & 2.39 & 0.14 & 2.39 & 0.13\\
\hline
\end{tabular}
\end{table}

\begin{table}
\begin{scriptsize}
\centering
\caption{ Distance Estimates for Sgr}
\label{DistTab}
\begin{tabular}{lcccccccc} \\ \hline
Name & $l$ & $b$ & $\rm(m-M)_0$ & D (kpc) & $\rm \sigma_D$ (kpc) & Reference & Method\\ \hline
MACHO & 5.0 & -8.0 & 16.97 & 24.8 & 0.8 & this paper & RRLy\\
MACHO & 5.0 & -4.0 & 16.71 & 22 & 1.0 & \citet{alc97} & RRLy\\
M54 & 5.6 & -14.1 & 17.19 & 27.4 & 1.5 & \citet{layden00} & 4 RRLy\\
M54 & 5-6.5 & -12 to -16 & 17.25 & 28.0 & 2.0 & Bellazzini, Ferraro \& Buonanno (1999) & 47Tuc\/HB stars \\
M54 & 5.6 & -14.1 & 17.10 & 26.3 & 1.8 & \citet{monaco04} & RGB Tip\\
M54 & 5.6 & -14.1 & 17.27 & 28.4 & 1.0 & \citet{siegel07} & isochrone\/MS fitting\\
3 Flds & 5.6 & -14.1 & 16.95 & 24.6 & 1.0 & \citet{marconi98} & HB\\
M54 & 5.6 & -14.1 & 17.02 & 25.4 & 1.0 & \citet{sarajedini95} & RHB-RGBC\\
M54 & 5.6 & -14.1 & 17.00 & 25.1 & 4.0 & \citet{dacosta95} & 4 globulars\\
M54 & 5.6 & -14.1 & $\sim$16.99 & $\sim$25 & -- & \citet{ibata94} & CMD\\
25deg2 & 3.0 & -7.0 & 16.90 & 24.0 & 2.0 & \citet{alard96} & RRLy \\
  & 9.0 & -23.0 & 17.20 & 27.6 & 1.3 & \citet{fahlman96} & CMD \\
  & 8.8 & -23.3 & 17.18 & 27.3 & 1.0 & \citet{mateo96} & RRab, CMD \\
  & 6.6 & -16.3 & 17.02 & 25.4 & 2.4 & \citet{mateo95} & RRab\\
ASA184 & 11 & -40 & $\sim$16.8 & $\sim$22 & -- & \citet{majewski99} & Red Clump \\
SA71 & -13 & -35 & $\sim$17.24 & $\sim$28 & -- & \citet{dinescu00} & HB \\
\hline
\end{tabular}
\end{scriptsize}
\end{table}

\clearpage

\clearpage

\clearpage

\clearpage

\clearpage

\end{document}